\documentclass{article}

\pdfoutput=1 
\usepackage{arxiv}

\usepackage[utf8]{inputenc} 
\usepackage[T1]{fontenc}    
\usepackage{hyperref}       
\usepackage{url}            
\usepackage{booktabs}       
\usepackage{amsfonts}       
\usepackage{nicefrac}       
\usepackage{microtype}      
\usepackage{lipsum}		
\usepackage{graphicx}
\usepackage[numbers]{natbib}
\usepackage{doi}

\usepackage{wrapfig}
\usepackage{xcolor,colortbl}

\definecolor{LightCyan}{rgb}{0.88,1,1}

\title{A Monotonicity Constrained Attention Module for Emotion Classification with Limited EEG Data}

\author{     \href{https://orcid.org/0000-0002-4862-7182}{\includegraphics[scale=0.06]{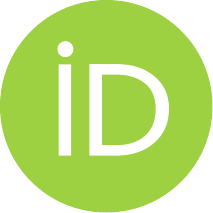}\hspace{1mm}Dongyang Kuang}\thanks{This preprint has not undergone peer review. The link to The Version of Record of this contribution can be found at https://link.springer.com/chapter/10.1007/978-3-031-16760-7\_21} \\
	School of Mathematics (Zhuhai)\\
	Sun Yat-sen University\\
	Zhuhai, Guangdong, China \\
	\texttt{dykuang@outlook.com} \\
	\And
	\href{https://orcid.org/0000-0002-6356-233X}{\includegraphics[scale=0.06]{orcid.pdf}\hspace{1mm}Craig Michoski} \\
	Oden Institute for Computational Engineering and Sciences\\
	University of Texas at Austin\\
	Austin, USA\\
	\texttt{michoski@oden.utexas.edu} \\
	 \AND
	 Wenting Li \\
	School of Mathematics (Zhuhai)\\
	Sun Yat-sen University\\
	Zhuhai, Guangdong, China \\
	 \texttt{liwt83@mail2.sysu.edu.cn} \\
	 \And
	 Rui Guo \\
	School of Mathematics (Zhuhai)\\
	Sun Yat-sen University\\
	Zhuhai, Guangdong, China \\
	 \texttt{guor36@mail2.sysu.edu.cn} \\
}

\date{}

\renewcommand{\shorttitle}{MCAM for emotion classification}

\begin{document}
\maketitle

\begin{abstract}
    In this work, a parameter-efficient attention module is presented for emotion classification using a limited, or relatively small, number of electroencephalogram (EEG) signals. This module is called the Monotonicity Constrained Attention Module (MCAM) due to its capability of incorporating priors on the monotonicity when converting features' Gram matrices into attention matrices for better feature refinement. Our experiments have shown that MCAM's effectiveness is comparable to state-of-the-art attention modules in boosting the backbone network's performance in prediction while requiring less parameters. Several accompanying sensitivity analyses on trained models' prediction concerning different attacks are also performed. These attacks include various frequency domain filtering levels and gradually morphing between samples associated with multiple labels. Our results can help better understand different modules' behaviour in prediction and can provide guidance in applications where data is limited and are with noises.
\end{abstract}


\section{Introduction}
\indent Due to improved computational methodologies alongside affordable access to efficient and powerful computational and neuroimaging hardware, there has been significant enthusiasm for the development of techniques for analyzing, predicting, and understanding human behavior through brain signals recorded from neuroimaging devices. One of the most popular and widespread neuroimaging techniques is electroencephalography (EEG), which is appealing for a variety of reasons, including that electroencephalograms can capture excellent time resolution as far as neuroimaging techniques go while being recorded on pragmatic devices that are portable, available, and affordable.

As stated in \cite{lotte2018review}, EEG classification algorithms can be roughly divided into the following five categories: i) conventional classifiers \cite{schlogl2009adaptive,li2010bilateral,liu2010improved,liu2012unsupervised}, ii) matrix and tensor based classifiers \cite{congedo2017riemannian}, iii) transfer learning based methods \cite{blankertz2008invariant,fazli2009subject}, iv) deep learning algorithms and advanced statistical approaches \cite{cecotti2010convolutional,lu2016deep}, and v) multi-label classifiers \cite{lotte2007review,blankertz2008invariant,steyrl2016random}.
Although many effective classification methods exist, particularly those utilizing deep learning techniques, many potential concerns remain for developing practical algorithms which can be deployed for general contextual use. Three such problems include:
{\color{blue}1)} large open-sourced EEG data sets are limited, making deep neural networks with a lot of parameters challenging to train and generalize effectively; 
{\color{blue}2)}  brain signals such as scalp EEG are known to have a high signal-to-noise ratio, effectively polluting the training and generalizability of the models \cite{leite2018deep,Bang2013NoiseRI,Repov2010DealingWN}; 
and {\color{blue}3)} the result from large and deep networks -- while accurately predictive --- can be challenging to interpret \cite{Cao2022TowardsID,Acharya2021EEGSC}. 

One solution to the above three concerns {\color{blue} (1)-(3)} is to construct parameter-efficient models that can be trained on relatively small and potentially noisy data sets while being lightweight enough to allow for physically/medically/clinically interpretable solutions. We note that good candidates for such approaches are attention modules in neural networks such as in \cite{hu2018squeeze,woo2018cbam,ash2017self-attention,dosovitskiy2020image}. Thus this work, inspired by the self-attention mechanism and Gram feature matrix in the context of neural style transfer \cite{gatys2016image}, presents the Monotonicity Constrained Attention Module (MCAM) that can dynamically construct the attention matrix from the feature Gram matrix. With MCAM, one can set constraints on different prior monotonic patterns to guide neural networks for selectively emphasizing informative features and suppressing unfavorable ones, leading to an efficient, accurate, and ultimately more easily interpretable framework.

\section{Related Work}

\noindent\textbf{Attention Mechanism}: Many attention mechanisms exist for refining deep features in the framework of neural network models. Among them, the Squeeze-and Excitation (SE) attention module \cite{hu2018squeeze} and Convolutional Block Attention Module (CBAM) \cite{woo2018cbam} are two representatives. The former helped win the last ImageNet contest in 2017. The latter performed attention operations both spatially and channel-wise. Both the two attention modules can be applied to any existing network. 
More recently, starting from the research done in \cite{dosovitskiy2020image} with attention matrix computed from query-key-value (QKV) feature branches, various types of self-attention mechanism are growing fast in different fields such as computer vision (CV), e.g.\cite{Lee2019SelfAttentionGP,Sankar2020DySATDN} and speech processing,e.g.\cite{Mittag2021NISQAAD,Koizumi2020SpeechEU}. 

\noindent\textbf{Our Work}: With the setting of limited and noisy data, our primary contributions can be summarized as follows: {\color{blue} \textit{1)}} We have developed a Monotonicity Constrained Attention Module (MCAM) suitable for EEG-based emotion classification when data is limited. Our experiments show that MCAM can help achieve performance comparable to other SOTA modules requiring fewer trainable parameters. {\color{blue} \textit{2)}} MCAM opens a portal in the backbone network so that one can conveniently incorporate priors on the monotonicity of the learned function that can effectively convert feature-based Gram matrices into attention matrices for better feature refinement. {\color{blue} \textit{3)}} For better interpretation, extra sensitivity analysis on MCAM's prediction concerning different attacks is also performed to investigate the various influences caused by inserting attention modules. 

\section{Proposed Method}

\begin{wrapfigure}{r}{0.3\textwidth}
	\includegraphics[width=0.3\textwidth, height=100pt, trim = 0 0 0 0, clip]{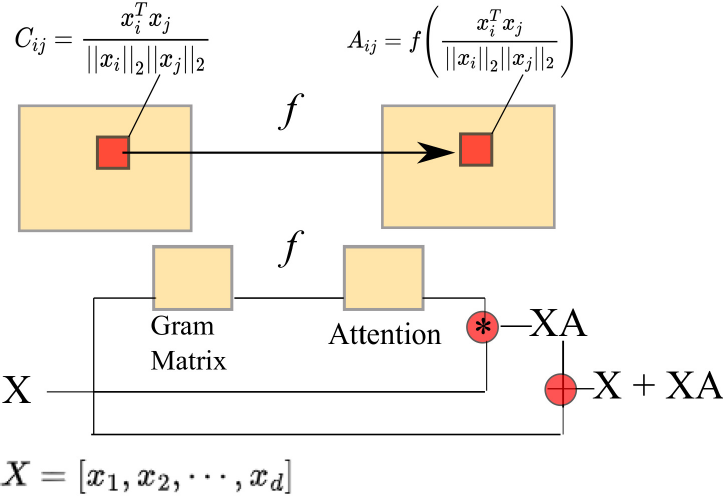}
	\caption{The Monotonicity Constrained Attention Module (MCAM).  Here $\|\cdot\|_{2}$ denotes the standard $L^2$-norm. }\label{fig:MCAM}
\end{wrapfigure} 

The mechanism of our proposed attention module is summarized in Figure \ref{fig:MCAM}. First, the Gram matrix $C$ \cite{sreeram1994properties} is computed using the deep channel features $X$. Next, a function $f: [-1,1] \rightarrow [0, 1]$ is constructed in the module, which is meant to `learn' the mapping that effectively translates $C$ element-wise to an attention matrix $A$ for better classification. As the key component of MCAM, we use a 3-layer MLP for approximating $f$ during training.

To understand the effect the trained function $f$ has for incorporating prior information between feature correlation and attention, we test different constraints for regularizing the monotonicity of $f$. In this manuscript, three configurations of MCAM are considered: 1) \textcolor{blue}{\textit{M1}}, no constraint on $f$'s monotonicity at all; 2) \textcolor{blue}{\textit{M2}}, $f$ should be non-decreasing on $[-1,0]$ and non-increasing on $[0,1]$, meaning the prior that less (positively or negatively) correlated features should contribute more to the corresponding value in $A$; and 3)  \textcolor{blue}{\textit{M3}}, $f$ should be non-increasing on $[-1,0]$ and non-decreasing on $[0,1]$ meaning the prior that the more correlation (positively or negatively) should result in more attention strength through $A$. 

The monotonicity constraint is defined using a uniform grid $\{t_i\}, i = 0,\cdots, N$, on $[-1, 1]$, where for for case \textcolor{blue}{\textit{M2}} the loss becomes: \begin{equation}
	\sum\limits_{i=0}^{N/2} \frac{1}{2}\left( |f'(t_i)| - f'(t_i)\right)) + \sum\limits_{i=N/2}^{N} \frac{1}{2}\left( |f'(t_i)| + f'(t_i)\right),
\end{equation}and for case \textcolor{blue}{\textit{M3}} becomes: \begin{equation}\sum\limits_{i=0}^{N/2} \frac{1}{2}\left( |f'(t_i)| + f'(t_i)\right)) + \sum\limits_{i=N/2}^{N} \frac{1}{2}\left( |f'(t_i)| - f'(t_i)\right).\end{equation}  Here $f'$ is estimated using a simple first order finite difference scheme. Finally, the resulting attention matrix $A$ incorporates into the deep feature refinement via the expression $X+XA$ as shown in Figure \ref{fig:MCAM}. 


\section{Experiments}

\noindent\textbf{EEGNet}: EEGNet is a compact convolutional neural network for EEG-based brain-computer interfaces which was proposed by \cite{ver2016EEGNet}.
We choose EEGNet as the backbone for benchmarking different attention modules for two main reasons: 
{\color{blue} 1)} It is one representative model that is parameter efficient, thus very suitable for small datasets;
{\color{blue} 2)} Its simple design allows for accessible examinations and interpretations on the effects of attention modules considered in this paper.

\noindent\textbf{Data Collection}: DEAP dataset\cite{5871728} is a well known database for benchmarking various emotion classification and analysis methods \cite{Khateeb2021MultiDomainFF,Staji2021EmotionRB,Kulkarni2021AnalysisOD,S2019AnalysisOE}. 
The dataset contains 32-channels EEG signals and 8-channels peripheral physiological signals from 32 volunteers who were asked to watch 40 1-min videos and report their emotion scores (varying from 1 to 9) in the four categories: valence, arousal, dominance, and liking. We will only be using the 32-channels EEG for experiments.

\noindent\textbf{Data Preprocessing}: 
We are interested in the within-subject classification task in this study. 
Setting the threshold at 5, we convert emotion scores from valence/arousal categories to form a 4-class classification family, comprised of $\{$HVHA, HVLA, LVHA, LVLA$\}$ for each subject. The first three seconds of each trial are baseline data and are used to normalize the rest via $x(t)\leftarrow x(t)/max(|x(t\leq3)|)$. 
Note that Subject 23 is excluded in the following experiments because this subject has only three emotion labels.


\noindent\textbf{Experiment Setting}:
For each subject, the data is split into three parts. The first 5000 time points (from  $\sim$0s to $\sim$39s) will be used in training, the following 5000 - 6000 (from $\sim$40s to $\sim$47s) will be used for validation,and the remaining segments (from $\sim$47s to $\sim$63s) are used for testing. 






\begin{wraptable}{l}{0.35\textwidth}
	\vskip-0.4cm
	\centering
	\caption{One sided paired T-tests on the F1-score, using the following abbreviations:. A: EEGNet, B: +CBAM, C: +SE,
		D: +\textcolor{blue}{\textit{M1}}, E: +\textcolor{blue}{\textit{M2}}, F: +\textcolor{blue}{\textit{M3}}, $\mu$: mean F1 score.}\label{tab:ttest}
	\begin{tabular}{c|c|c}
		\hline
		$H_0$ &  $H_1$  & p-value\\
		\hline
		$\mu_A = \mu_F$ &  $\mu_A < \mu_F$ & 0.018\\
		$\mu_B = \mu_F$ &  $\mu_B < \mu_F$ & 0.038\\
		$\mu_F = \mu_C$ &  $\mu_F < \mu_C$ & 0.115\\
		$\mu_D = \mu_F$ &  $\mu_D < \mu_F$ & 0.437\\
		\hline
	\end{tabular}
\end{wraptable}

Note that in order to mitigate the problems associated to limited data and imbalanced labeling, a data generator is used to randomly crop segments of 1s to provide batches for feeding the network during training and validation.  Within each batch, the generator guarantees that each label is associated with about 25\% of the total samples generated. For a valid and consistent comparison among different models, the test set will be cropped into non-overlapping segments of 1s. For each subject and each attention module compared, training is repeated 10 times. During each training repetition, the best model in terms of validation accuracy is reloaded to make predictions on the testing set, and one instance of the performance under that configuration is stored. For all experiments, the following hyperparameters are used: the backbone EEGNet's dropout rate is set to 0.5, the batchsize is set to 256, and the optimization is performed with an Adam optimizer using a learning rate of $10^{-3}$. For training with the proposed MCAM, the weight for the extra loss constraining monotonicity is set at 0.1. These hyperparameters were selected from our pre-experiments with a small amount of data. Same hyperparamters were used for all experiments. The code used for our experiments will be made available at \url{https://github.com/dykuang/BCI-Attention}





\begin{table}[htbp]
	\centering
	\caption{Mean (±std) performance with different attention modules inserted. }\label{tab:benchmark}
	\begin{tabular}{cccccccc}
		\hline
		Method & EEGNet & +QKV& +CBAM & +SE & +\textcolor{blue}{\textit{M1}} & +\textcolor{blue}{\textit{M2}} & +\textcolor{blue}{\textit{M3}} \\
		\hline
		Params & 3020 & 4109 & 3200 & 3102 & \textbf{3061} & \textbf{3061} & \textbf{3061} \\
		\hline
		Acc.(\%) & 93.9{\tiny±5.4} & 93.5{\tiny±5.0}&94.3{\tiny±4.7} & 95.4{\tiny±3.8} & 95.0{\tiny±4.7} & 93.9{\tiny±6.1} & 95.0{\tiny±4.2}\\
		Spec.(\%) & 97.7{\tiny±1.9} & 97.4{\tiny±2.3}&97.9{\tiny±1.7} & 98.2{\tiny±1.5} & 98.0{\tiny±1.9} & 97.5{\tiny±2.6} & 97.9{\tiny±2.0}\\
		F1(\%) & 94.0±{\tiny5.4} & 93.5{\tiny±5.0}&94.3{\tiny±4.6} & 95.4{\tiny±3.8} & 94.9{\tiny±4.7} & 93.8{\tiny±6.2} & 95.0{\tiny±4.2}\\
		\hline
	\end{tabular}
\end{table}


\noindent\textbf{Benchmark}: We summarize commonly used classification metrics in Table \ref{tab:benchmark}, using EEGNet as the backbone, and inserting all attention modules at the same location in our benchmark. 
Notice that the QKV type self-attention performs the worst, demonstrating no improvement. We hypothesize this is due to limited training data, as pointed out in \cite{dosovitskiy2020image}. For the remainder of this section, we focus only on the attention modules where performance is equivalent to, or higher than the backbone model. For a more quantitative comparison, we also perform paired T-tests and collect the resulting p-values in Table \ref{tab:ttest}.  At a significance level of $\alpha = 0.02$ the alternative hypothesis $H_1$ is accepted, i.e. EEGNet+\textcolor{blue}{\textit{M3}} has a higher F1-score than EEGNet alone. At significance level of $\alpha = 0.05$, EEGNet+\textcolor{blue}{\textit{M3}} has a higher F1-score than EEGNet+CBAM. Note that there is not enough evidence (at the significance level of 0.05) to reject the null hypothesis for the rest tests. That is, EEGNet$+$\textcolor{blue}{\textit{M3}} performs similarly to EEGNet$+$SE and EEGNet$+$\textcolor{blue}{\textit{M1}}. 
\begin{figure}[htbp]
	\centering
	\includegraphics[width=0.9\textwidth, height=140pt, trim=60 0 60 0, clip]{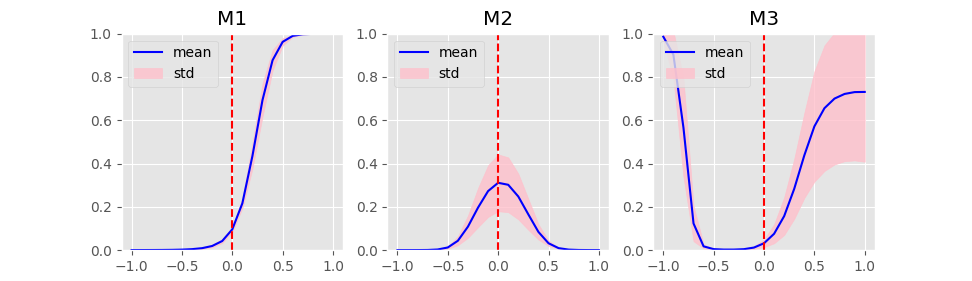}
	\caption{Mapping from the Gram matrix's entries $C_{ij}$ to the attention matrix's entries $A_{ij}$ by MCAM. This plot was made from the 10 models trained from Subject 24's data.}
	\label{fig:mono_trace}
\end{figure}

Finally we can check the different monotonic patterns learned for different subjects during training. Figure \ref{fig:mono_trace} shows one example of the monotonic pattern mapping from $C_{ij}$ to $A_{ij}$ learned with different prior constraints. 

\subsection{Models' scalp attention pattern}
\begin{figure}[htbp]
	\centering
	\includegraphics[width=0.9\textwidth, height=140pt, trim=0 0 10 0, clip]{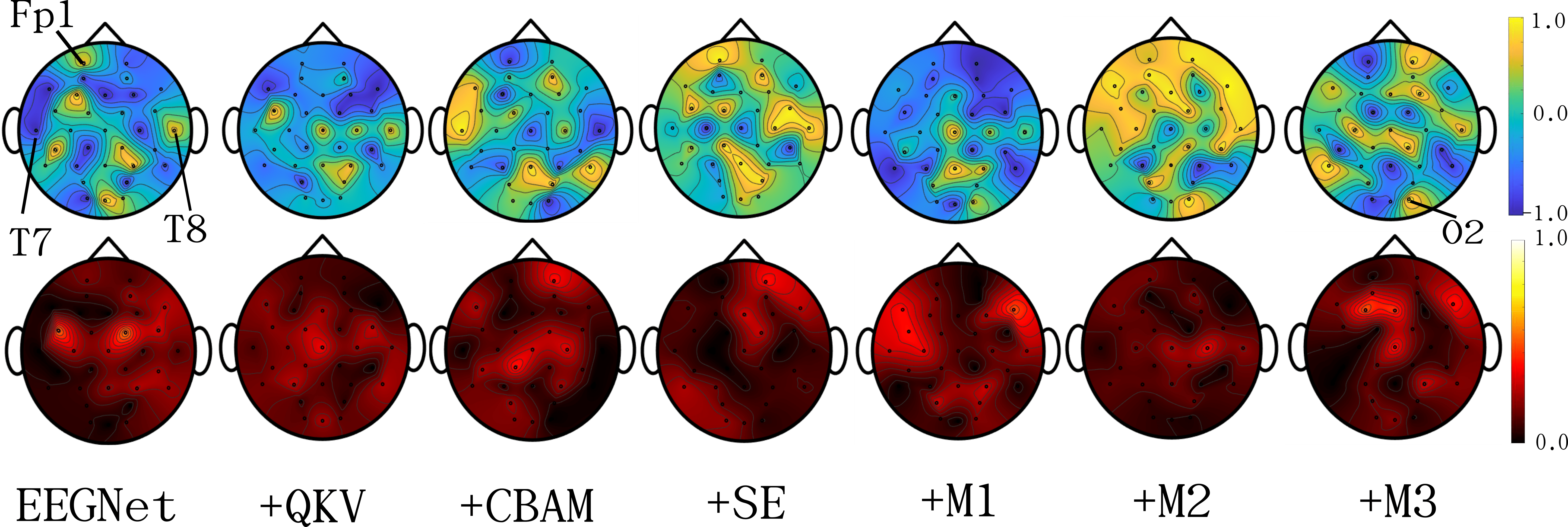}
	\caption{Scalp map with backbone EEGNet's kernel weights for spatial attention pattern in the first depthwise convolution layer. First row: mean values calculated from 10 trained models. Second row: the standard deviation associated with the result from the first row.}
	\label{fig:scalp}
\end{figure}
The insertion of different attention modules can potentially change the spatial attention pattern of the backbone EEGNet. For comparison, we can visualize the scalp map with EEGNet's kernel weights (normalized to $[-1, 1]$) for its spatial attention pattern, i.e., kernel weights in the first depthwise convolution layer. As one can observe from Figure \ref{fig:scalp}, global patterns vary from model to model, though they share the same backbone network. Locally speaking, except for the case +\textcolor{blue}{\textit{M2}}, the value pair learned for channels T8 and T7 have different signs, while the case +CBAM learned the opposite pattern compared to the rest. With the exception of +\textcolor{blue}{\textit{M1}}, all other configurations show a locally isolated island in the Fp1 area. All three variations of MCAM considered here also show a similar attention pattern around the O2 area. The standard deviation shown can be interpreted loosely as corresponding to each model's confidence in the value of its coincident kernel weights. Models compared here show high confidence in most areas, where areas of low confidence offer an interesting opportunity for deeper analysis and potentially clinical interpretation. It is also worth noting that, similar to the mean value patterns, the global patterns for standard deviation appear quite different across different models. Whether or not these patterns can be tied to clinical findings remains a question. On the other hand, robust algorithms that can help encode one's prior knowledge about clinical patterns into the models' weights is also an import research direction.

\subsection{Models' sensitivity of prediction on inputs' frequency}
\begin{figure}[tbh]
	\centering
	\includegraphics[scale=0.40, trim=20 0 40 0, clip]{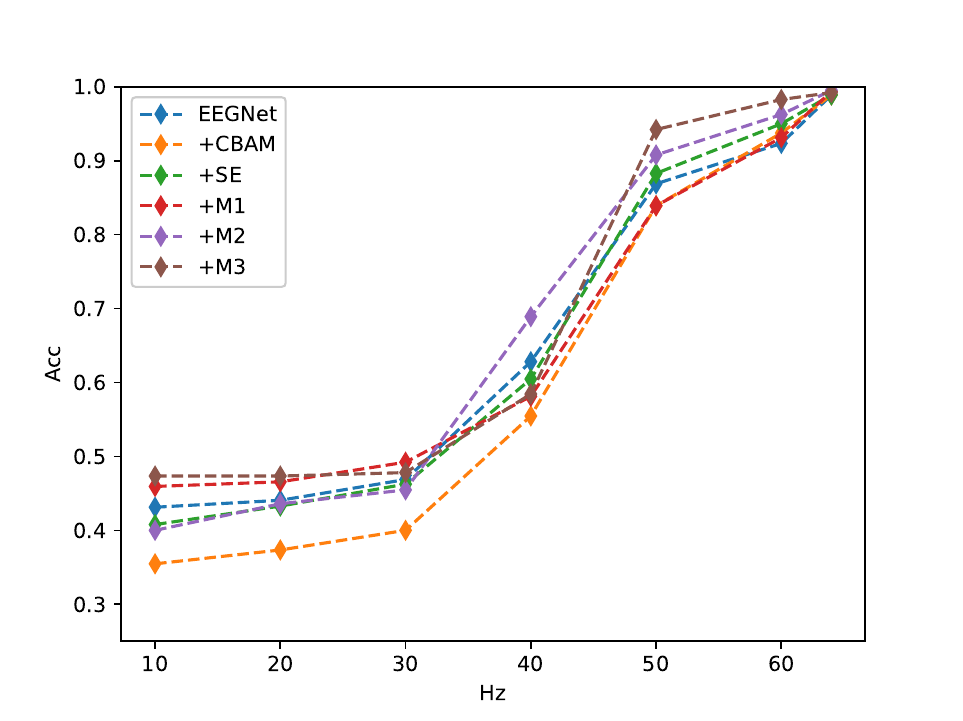}
	\includegraphics[scale=0.40, trim=20 0 40 0, clip]{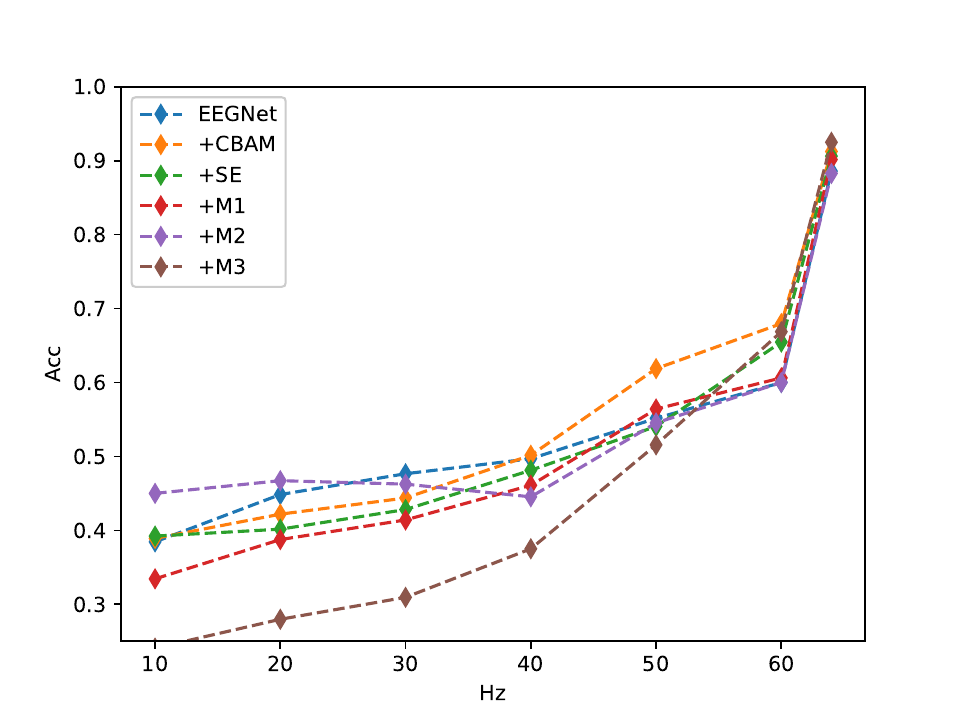}
	\caption{The track of performance when inputs are lowpass filtered to different frequencies. Left: Subject 12, Right: Subject 24. The frequency used for the plot are 10Hz, 20Hz, 30Hz, 40Hz, 50Hz, 60Hz and 64Hz.}
	\label{fig:freq_trace}
\end{figure}

Low pass filters are frequently used in applications to suppress noise as a preprocessing step. This section examines how frequency information in the input affects the trained model predictions with different modules inserted. With the same test data, first, a lowpass filter is used at different frequencies, and then the accuracy of different model predictions is tracked on this filtered input. Figure \ref{fig:freq_trace} demonstrates the result for subject 12 and subject 24. Of note is that different models show the same trend given the same data, but the trends vary across different subjects. For subject 12, a noticeable drop in performance is observed when inputs are lowpass filtered below 50Hz, while stable performance is recovered when the frequencies are set below 30Hz. For subject 24, a noticeable drop in performance is seen at 60Hz, while the performance trace decreases slowly as frequencies go lower. In the case of subject 24, suppressing high-frequency noise using lowpass filters seems to compromise model performance. While each model decreases in performance as higher frequencies are filtered out, the configuration of EEGNet+\textcolor{blue}{\textit{M2}} has an opposite trend in the range of $20\sim 40$Hz. This observation raises what seems to serve as a cautionary tale that it is crucial to take care when smoothing data in the frequency domain, as the response may be complicated and hard to predict a priori.
\subsection{Model sensitivity on morphisms between samples.}
\begin{figure}[tbh]
	\centering
	\includegraphics[width=0.75\textwidth, height=280pt, trim=0 0 0 0, clip]{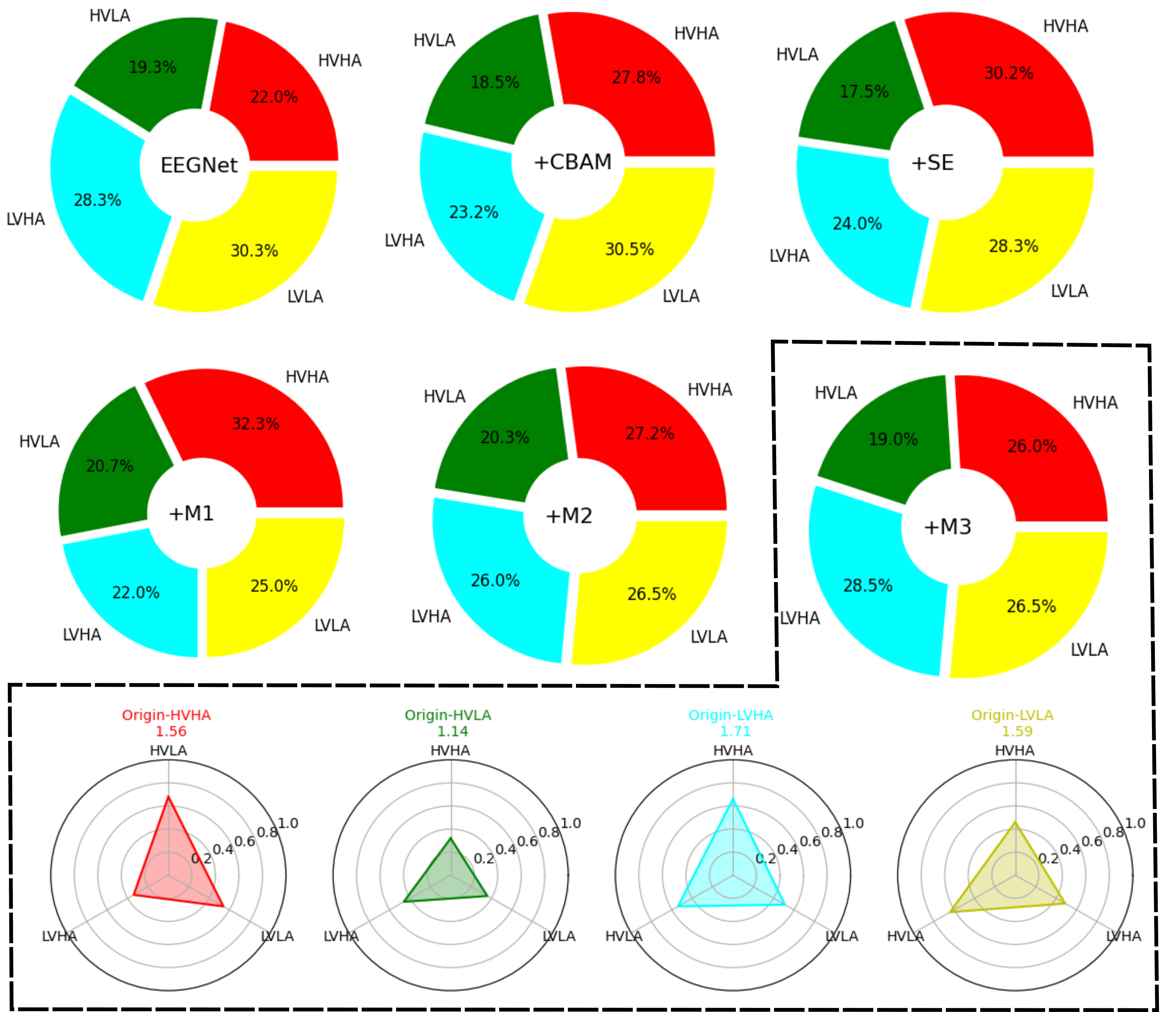}
	\caption{Visualization of how different models tend to change prediction labels under the linearly interpolating morphism. The samples for making the plot were taken from subject 24. The colored number shown are the actual number of $S_i$. For example, 1.56/6 = 0.26 recovers the percentages shown in the donut plot. }
	\label{fig:PTR}
\end{figure}
Considering two samples $x_i$ and $x_j$ each associated with a different label $i$ and $j$ in the $N$ classification problem, we define a morphism $g_i^j$ parameterized by $u \in [0,1]$ such that $x_i=g_i^j(0)*x_i$ and $x_j=g_i^j(1)*x_i$. The abstract operation $*$ will be made explicit below, as an example. 
We further note that in the above setting, a value $u_i^j$ always exists such that the model's confidence score (usually represented by the softmax value from the last dense layer's output) for label $i$ first drops below its score for label $j$. The lower the value of $u_i^j$ is, the more likely the model under examination will change its prediction on sample $x_i$ from label $i$ to label $j$. 

For visualization, we set $u_i^j$ as a point in polar coordinates $(\rho, \theta) = (u_i^j, \frac{2\pi v_j}{N-1})$ where $v_j = 0,\cdots,N-2$ is some indexing for different $j$. These points will serve as the vertices of an $N-1$ polygon. The resulting summation $S_i = \sum_{j\neq i} u_i^j$ then represent how likely a model is to choose label $i$ against all other labels under the considered morphism $g$. Furthermore, if $u_i^j + u_j^i = 1$ then $\sum_{i=0}^{N-1} S_i= \frac{N(N-1)}{2}$, in which case we can check the values of $S_i, i=0,\cdots, N-1$ among different models on the same selected ``representative'' samples per label for comparing their "preferences" among candidate labels under the chosen morphism $g$.

As a demonstration of the above concept, we choose the simple discrete linear interpolation for $g$, defined by $g_i^j(u)*x_i = (1-u)x_i+ux_j$ as the morphism operation, where samples $\{x_i\}$ are selected so that all considered models have correct predictions on them. The resulting summaries are shown in Figure \ref{fig:PTR}. Each donut plot of the above two rows is the visualization of $S_i$ for a specific model. The last row gives an example of expanding more detailed information per slice by visualizing a triangle (3-polygon) for each label.  All models being compared here assign the lowest values to the HVLA category, meaning models are more likely to assign other labels for the input HVLA sample when $g$ morphs it away to other samples of different labels. Also, note that both the backbone model and the case when CBAM is inserted have the largest value for LVLA. The insertion of the SE module, MCAM( \textcolor{blue}{\textit{M1}}) and MCAM( \textcolor{blue}{\textit{M2}}) assign the highest value to the HVHA category. Of further note is that the two configurations of MCAM ( \textcolor{blue}{\textit{M2}} and  \textcolor{blue}{\textit{M3}}) are surprisingly similar considering that their monotonicity mapping in the Gram matrix C's element to attention matrix A are opposite (See figure \ref{fig:mono_trace}). Additionally, we examine how the scores $S_i$ for $i=0,1,2,3$ are spread across the four labels by checking the standard deviation (std), thus indicating how each model distributes their prediction "preference" among the 4 categories given the selected input samples and morphism $g$.  Here we find that the std for the backbone EEGNet is 0.269, while the insertion of CBAM and SE raises the std to 0.274 and 0.292, respectively, where the distributions are notably more skewed.  However, MCAM( \textcolor{blue}{\textit{M1}}) has an std of 0.271, which is the close to the case where no attention module is inserted at all, while MCAM( \textcolor{blue}{\textit{M2}}) has by far the lowest value std of 0.164 followed by a std of 0.215 for the MCAM( \textcolor{blue}{\textit{M3}}).

\section{Conclusion}
We have constructed a parameter-efficient attention module called MCAM for emotion classification tasks with limited EEG data. MCAM allows one to put constraints on the embedded function's monotonicity for mapping a deep feature Gram matrix to form an effective attention matrix during feature refinements. Experiments show that MCAM's effectiveness is comparable to state-of-the-art attention modules with additional benefits. Additionally, a sensitivity analysis with different levels of lowpass filtering has been conducted, along with a novel morphing analysis designed to improve insight into the model's behavior through visualization. 
Future work will focus on designing more generalizable and interpretable models on limited and corrupted data sets.  
\section{Acknowledgement}
\textit{This work was partially supported by the Fundamental Research Funds for the Central Universities, Sun Yat-sen University (22qntd2901)}

\bibliographystyle{abbrvnat}
\bibliography{MILLan_ref}

\end{document}